\documentclass[a4paper,11pt]{article}
\usepackage{pos}

\newcommand{\beqn}{\begin{eqnarray}}
\newcommand{\eeqn}{\end{eqnarray}}
\newcommand{\be}{\begin{equation}}
\newcommand{\ee}{\end{equation}}

\newcommand{\ba}{\begin{array}{c}}
\newcommand{\bat}{\begin{array}{cc}}
\newcommand{\ea}{\end{array}}

\newcommand{\bi}{\begin{itemize}}
\newcommand{\ei}{\end{itemize}}

\newcommand{\ket}{\,\rangle}
\newcommand{\bra}{\langle \,}

\newcommand{\Frac}[2]{\frac{\displaystyle #1}{\displaystyle #2}}

\newcommand{\cO}{{\cal O}}

\newcommand{\mF}{\mathcal{F}}

\newcommand{\gsim}{\stackrel{>}{_\sim}}

\newcommand{\comment}[1]{}

\title{Heavy states and electroweak effective approaches}

\author*[a,\dag]{Ignasi Rosell}
\notes{\note{We wish to thank the organizers for the pleasant conference. This work has been supported in part by the Spanish Government and ERDF funds from the European Commission (FPA2016-75654-C2-1-P, FPA2017-84445-P, PID2019-108655GB-I00); by the Generalitat Valenciana (PROMETEO/2017/053); by the Universidad Cardenal Herrera-CEU (INDI20/13); by the EU STRONG-2020 project under the program H2020-INFRAIA-2018-1 [grant agreement no. 824093]; and by the STSM Grant from COST Action CA16108.
}}
\author[b]{Antonio Pich}
\author[c]{Juan Jos\'e Sanz-Cillero}

\affiliation[a]{Departamento de Matem\'aticas, F\'\i sica y Ciencias Tecnol\' ogicas, Universidad Cardenal Herrera-CEU, CEU Universities, 46115 Alfara del Patriarca, Val\`encia, Spain}
\affiliation[b]{IFIC, Universitat de Val\`encia -- CSIC, Apt. Correus 22085, 46071 Val\`encia, Spain}
\affiliation[c]{Departamento de F\'\i sica Te\'orica and Instituto de F\'\i sica  de  Part\'\i culas  y  del  Cosmos IPARCOS,  Universidad Complutense de Madrid, E-28040 Madrid, Spain}

\emailAdd{rosell@uchceu.es}
\emailAdd{pich@ific.uv.es}
\emailAdd{jjsanzcillero@ucm.es}

\abstract{The existence of a mass gap between the Standard Model (SM) and possible new states encourages us to use effective field theories. Here we follow the non-linear realization of the electroweak symmetry breaking: the electroweak effective theory (EWET), also known as Higgs effective field theory (HEFT) or electroweak chiral Lagrangian (EWChL). At short distances an effective resonance Lagrangian which couples the SM states to bosonic and fermionic resonances is considered. After integrating out the resonances and assuming a well-behaved high-energy behavior, we estimate or bound purely bosonic low-energy constants in terms of only resonance masses. Current experimental information on these low-energy constants allows us to constrain the high-energy resonance masses.}

\FullConference{%
  *** The European Physical Society Conference on High Energy Physics (EPS-HEP2021), ***\\
  *** 26-30 July 2021 ***\\
  *** Online conference, jointly organized by Universität Hamburg and the research center DESY ***
}

\begin{document}
\maketitle

\section{Introduction}

The success of the Standard Model (SM) has been confirmed by the LHC and it implies the existence of a mass gap between SM fields and possible new-physics (NP) states. Therefore, effective field theories (EFTs) are very convenient in order to study systematically low-energy data to search for traces of NP scales, the so-called bottom-up approach. Whereas the low-energy constants (LECs) of these EFTs are free-parameters and contain information about heavy scales, the structure of the effective Lagrangian relies on the particle content (SM particles here), the symmetries and the power counting. The symmetries and the power counting depend on the mechanism to introduce the Higgs. Here we follow the most general case,  the electroweak (EW) effective theory (EWET): the non-linear realization of the electroweak symmetry breaking is taken and, consequently, no assumptions are made about the relation between the Higgs and the EW Goldstones and a ``chiral'' (generalized momenta) expansion is observed. It is worth noting that the so-called SM effective field theory (SMEFT) is a particular case of the EWET. At higher energies possible heavy resonances are included by following the same scheme, {\it i.e.}, same symmetries and power expansion.  

Accordingly, we work with two EFTs: the EWET at long distances, which includes only the SM fields, and an EW resonance theory at shorter distances, which contains in addition heavy resonances. After integrating out the resonances both Lagrangians can be matched and the EWET LECs are determined in terms of resonance parameters. Assuming  a good high-energy behavior in the resonance theory allows us to imagine this effective Lagrangian as a bridge between the EWET and the unknown underlying theory and also to reduce the number of unknown resonance parameters. Note that the main focus of this work is to constrain NP scales by using current experimental bounds on EWET LECs and our determinations of these LECs in terms of only a few resonance parameters~\cite{PRD,Pich:2015kwa}.

\section{Theoretical framework}

The EWET Lagrangian is organized following an expansion in generalized momenta~\cite{Weinberg,Buchalla,lagrangian}:
\begin{eqnarray}
\mathcal{L}_{\mathrm{EWET}} &=& \sum_{\hat d\ge 2}\, \mathcal{L}_{\mathrm{EWET}}^{(\hat d)}\,, \label{EWET-Lagrangian0}
\end{eqnarray}
where the chiral dimension $\hat d$ indicates the infrared behavior at low momenta~\cite{Weinberg}. The building blocks and their related power-counting rules can be found in Refs.~\cite{PRD,lagrangian}. In view of the current experimental information, the relevant bosonic part of the LO EWET Lagrangian reads
\begin{eqnarray}
\Delta \mathcal{L}_{\mathrm{EWET}}^{(2)} & =&  \frac{v^2}{4}\,\left( 1 +\frac{2\,\kappa_W}{v} h + \frac{c_{2V}}{v^2} \,h^2\right) \bra u_\mu u^\mu\ket    \, , \label{LOEWET_lagrangian}
\end{eqnarray}
where $h$ is the Higgs field, $u_\mu$ the tensor containing one covariant derivative of the EW Goldstones and $\langle\cdots\rangle$ indicate an $SU(2)$ trace. $\kappa_W$ and $c_{2V}$ parametrize the $hWW$ and $hhWW$ couplings, respectively. Taking into account again the available experimental information, the relevant part of the purely bosonic NLO EWET Lagrangian is given by~\cite{lagrangian}:
\begin{equation}
\Delta \mathcal{L}_{\mathrm{EWET}}^{(4)}  =
  \Frac{\mF_1}{4}\,\bra {f}_+^{\mu\nu} {f}_{+\, \mu\nu}- {f}_-^{\mu\nu} {f}_{-\, \mu\nu}\ket +
  \Frac{i\, \mF_3 }{2} \,\bra {f}_+^{\mu\nu} [u_\mu, u_\nu] \ket +
 \mF_4 \bra u_\mu u_\nu\ket \, \bra u^\mu u^\nu\ket +
 \mF_5 \bra u_\mu u^\mu\ket\, \bra u_\nu u^\nu\ket ,
 \label{EWET_lagrangian}
\end{equation}
where $f^{\mu\nu}_\pm$ introduce the field-strength tensors. In the SM, $\kappa_W=c_{2V}=1$ and $\mathcal{F}_{1,3,4,5}=0$. $\mF_1$, $\mF_{1,3}$ and $\mF_{1,3-5}$ contribute, respectively, to the $S$ oblique parameter, the trilinear and the quartic gauge couplings.

Since we are interested only in the resonance contributions to the bosonic $\cO(p^4)$ EWET LECs, only $\cO(p^2)$ operators with up to one bosonic resonance $R$ are required~\cite{lagrangian}. The relevant $CP$-even resonance Lagrangian reads~\cite{PRD,lagrangian}:
\begin{align}
\qquad \Delta \mathcal{L}_{\mathrm{RT}} \,=\,&
 \frac{v^2}{4}\!\left(\! 1 \!+\!\frac{2\,\kappa_W}{v} h \!+\! c_{2V} \,h^2\!\right)\! \bra u_\mu u^\mu\ket
  +\bra V_{\mu\nu} \left( \frac{F_V}{2\sqrt{2}}  f_+^{\mu\nu} + \frac{i G_V}{2\sqrt{2}} [u^\mu, u^\nu]  + \frac{\widetilde{F}_V }{2\sqrt{2}} f_-^{\mu\nu}   \right) \ket 
    \phantom{\bigg( }
 \nonumber \\
& \quad
+ \bra A_{\,\mu\nu} \left(\frac{F_A}{2\sqrt{2}}  f_-^{\mu\nu}  +  \frac{\widetilde{F}_A}{2\sqrt{2}} f_+^{\mu\nu} +  \frac{i \widetilde{G}_A}{2\sqrt{2}} [u^{\mu}, u^{\nu}]    \right) \ket  
 + \frac{c_{d}}{\sqrt{2}}\, S_1\bra u_\mu u^\mu \ket  \,.
 \label{Lagrangian}
\end{align} 
By integrating out the resonance fields in (\ref{Lagrangian}), the EWET Lagrangian of (\ref{EWET_lagrangian}) is recovered and, consequently, the EWET LECs are determined in terms of resonance parameters~\cite{PRD,lagrangian},
\begin{align}
\mF_1 &= - \Frac{F_V^2-\widetilde{F}_V^2}{4M_{V}^2} + \Frac{F_A^2-\widetilde{F}_A^2}{4M_{A}^2}\,, &
\mF_3 &= -  \Frac{F_VG_V}{2M_{V}^2} - \Frac{\widetilde{F}_A\widetilde{G}_A}{2M_{A}^2}\,, \nonumber \\ 
\mF_4 &= \Frac{G_V^2}{4M_{V}^2} + \Frac{{\widetilde{G}_A}^2}{4M_{A}^2} \,,&
\mF_5&= \Frac{c_{d}^2}{4M_{S_1}^2}-\Frac{G_V^2}{4M_{V}^2} - \Frac{{\widetilde{G}_A}^2}{4M_{A}^2}\,. \label{integration}
\end{align}

\begin{figure*}[!t]
\begin{center}
\begin{minipage}[c]{7.0cm}
\includegraphics[width=7.0cm]{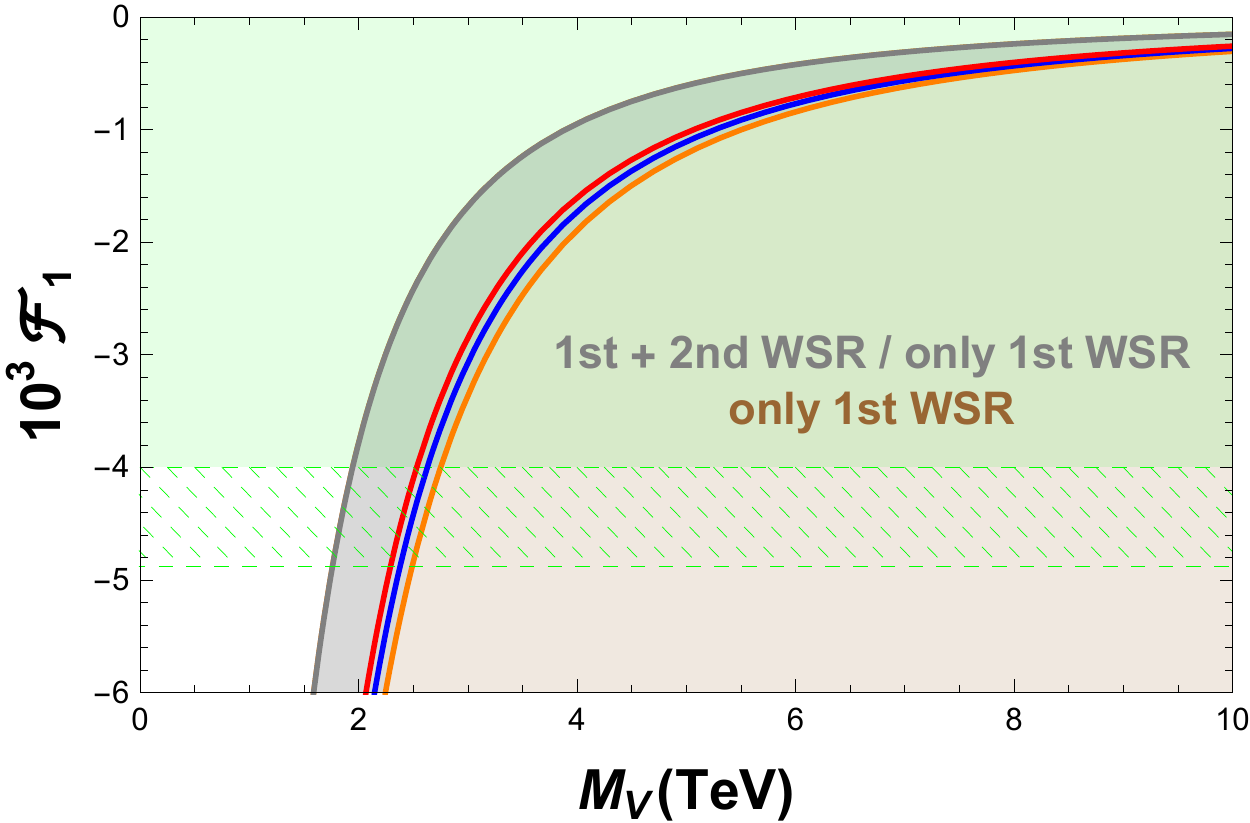}  
\end{minipage}
\hskip .5cm
\begin{minipage}[c]{7.0cm}
\includegraphics[width=7.0cm]{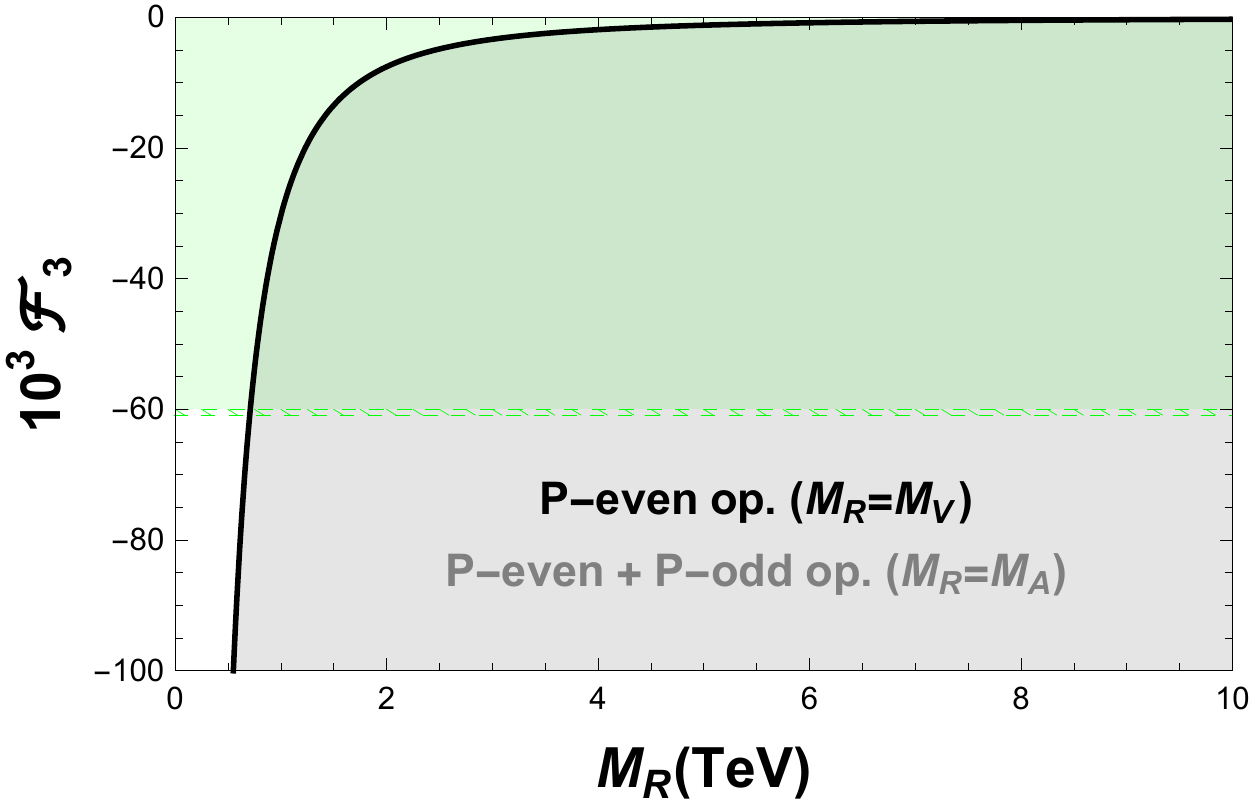} 
\end{minipage}
\\[8pt]
\begin{minipage}[c]{7.0cm}
\includegraphics[width=7.0cm]{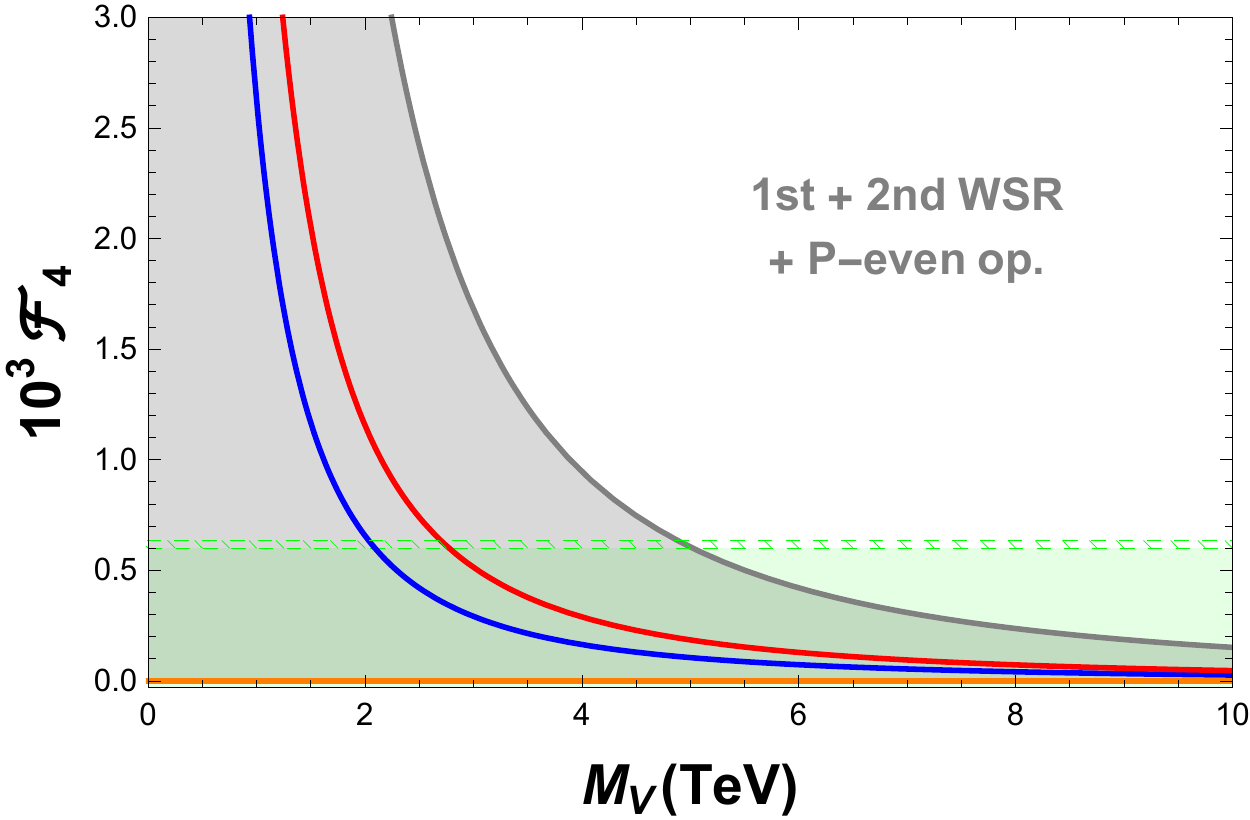} 
\end{minipage}
\hskip .2cm
\begin{minipage}[c]{7.0cm}
\includegraphics[width=7.0cm]{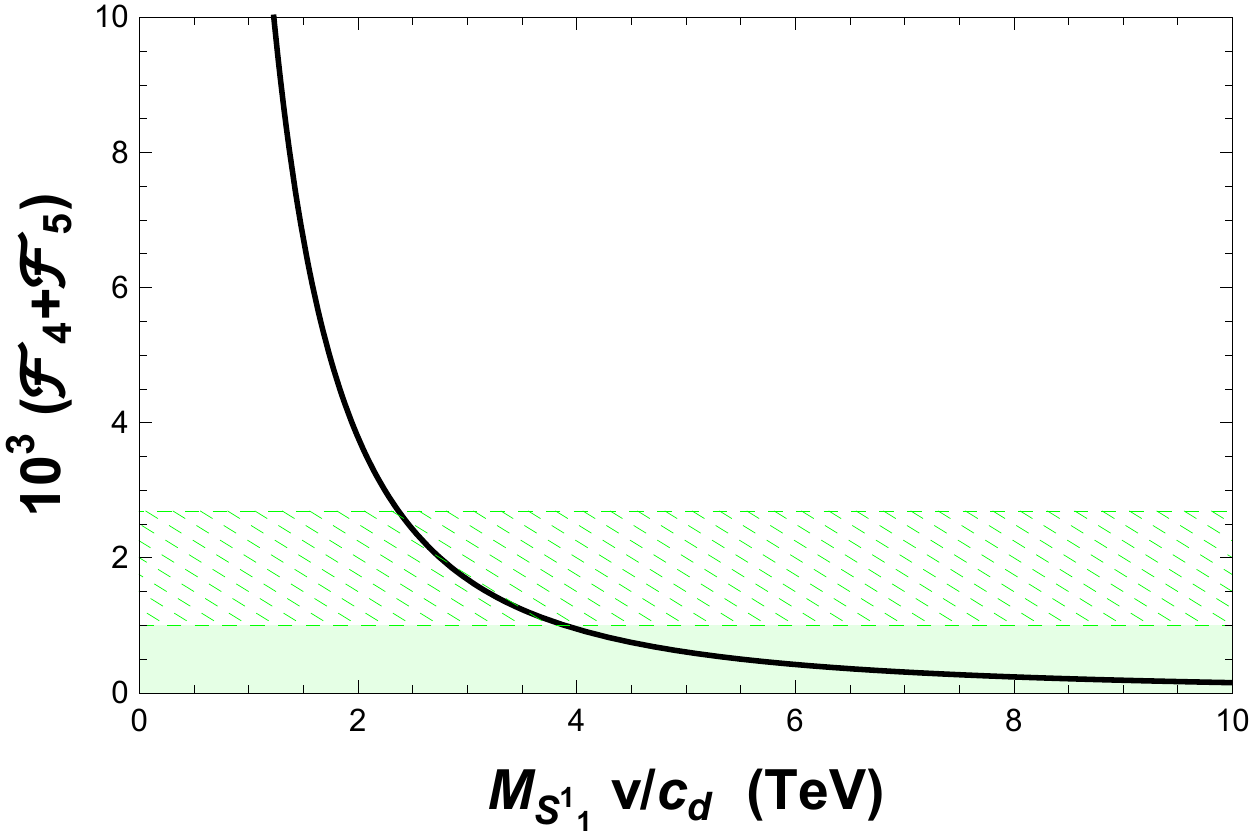}
\end{minipage}
\end{center}
\caption{{\small $\mF_1$, $\mF_3$, $\mF_4$ and $\mF_4+\mF_5$ as functions of the corresponding resonance mass ($M_V$, $M_A$ or $M_{S_1^1} \,v/c_d$)~\cite{PRD}.
The 95\% CL experimentally-allowed regions given in Table~\ref{exp} are covered by the green areas, and they are further extended by dashed green bands accounting for our estimation of the one-loop running uncertainties. If the prediction of the LECs depend on both $M_V$ and $M_A$, the gray and/or brown regions cover all possible values for $M_A>M_V$. If the 2nd WSR has been assumed, we indicate it explicitly in the plot, with the corresponding lines for the limit $M_A=M_V$ (orange), $M_A=1.1\,M_V$ (blue), $M_A=1.2\,M_V$ (red) and $M_A\to \infty$ (dark gray). In the case without the 2nd WSR, the prediction for $\mF_1$ is given by the gray and brown regions. In case of using only the even-parity operators, it is indicated.
}} 
\label{plots1}
\end{figure*}

As stressed previously, high-energy constraints allow us to reduce the number of resonance parameters. For the case at hand, the following short-distance constraints have been used~\cite{PRD,lagrangian}:
\begin{enumerate}
\item Well-behaved two-Goldstone vector form factor (VFF). Assuming that this form factor vanishes at high energies, the following constraint is found:
\begin{equation}
v^2\, -\,F_V\,G_V\,   -\,\widetilde{F}_A\,\widetilde{G}_A \,=\, 0 \,. \label{VFF}
\end{equation}
\item Weinberg Sum Rules (WSRs). The $W^3B$ correlator is an order parameter of the EWSB. In asymptotically-free gauge theories it vanishes at short distances as $1/s^3$, implying two superconvergent sum rules~\cite{WSR}: the 1st WSR (vanishing of the $1/s$ term) and the 2nd WSR (vanishing of the $1/s^2$ term). While the 1st WSR is supposed to be also fulfilled in gauge theories with nontrivial ultraviolet (UV) fixed points, the validity of the 2nd WSR depends on the particular type of UV theory considered~\cite{1stWSR}. The 1st and 2nd WSRs read, respectively,
\begin{equation}
F_V^2 +\widetilde{F}_A^2 - F_A^2 - \widetilde{F}_V^2 \,=\, v^2  \,, \qquad \quad
 F_V^2 M_V^2 + \widetilde{F}_A^2 M_A^2 - F_A^2 M_A^2 - \widetilde{F}_V^2 M_V^2=0 . \label{WSRs}
\end{equation}
\end{enumerate}
The use of (\ref{VFF}) and (\ref{WSRs}) in (\ref{integration}) lets us get the following determinations or bounds, which are displayed in the plots of Figure~\ref{plots1}~\cite{PRD}:
\begin{enumerate}
\item $\mF_1$. Assuming both WSRs, $\mF_1$ can be determined in terms of only resonance masses; contrary, discarding the 2nd WSR, we only find a bound in terms of only resonance masses~\cite{PRD}:\footnote{For this bound one needs to assume that $M_A>M_V$ and, in the case of considering $P$-even and also $P$-odd operators, $F_A^2>\widetilde F_A^2$. Both hypothesis seem to be reasonable working assumptions.}
\begin{equation}
\mF_1\Big|^{\mathrm{1st\,\&\,2nd\,WSRs}} \,=\, -\displaystyle\frac{v^2}{4} \left( \frac{1}{M_V^2} \!+\! \frac{1}{M_A^2} \right)\,, \qquad  \qquad
\mF_1\Big|^{\mathrm{1st\,WSR}}\,<\,  \displaystyle\frac{-v^2}{4M_V^2}\,. \label{F1}
\end{equation}
These results are shown in the top-left plot in Figure~\ref{plots1}. The dark gray curve shows the upper bound of (\ref{F1}). Therefore, and if only the 1st WSR is obeyed, the whole region below this line (gray and brown areas) would be theoretically allowed. If one accepts moreover the validity of the 2nd WSR, $\mF_1$ is determined as a function of $M_V$ and $M_A$, see (\ref{F1}), being the dark gray curve the limit $M_A\to\infty$. The values of $\mF_1$ for some representative axial-vector masses are shown in the red ($M_A=1.2\, M_V$), blue ($M_A=1.1\, M_V$) and orange ($M_A=M_V$) curves; being the orange line the lower bound $\mF_1=-v^2/(2M_V^2)$. This range of $M_A\sim M_V$ corresponds actually to the most plausible scenario~\cite{ST}. Then, and if both WSRs are followed, only the gray region  would be theoretically allowed if one assumes that $M_A>M_V$.\footnote{In the case with only $P$-even operators, following both WSRs implies $M_A>M_V$, but this is only a reasonable assumption if one considers also $P$-odd operators.}
\item $\mF_3$. In this case the WSRs are not relevant. Assuming only $P$-even operators, (\ref{VFF}) implies a determination in terms of only resonance masses, whereas the inclusion of $P$-odd operators reduces this determination to a bound~\cite{PRD}:\footnote{For this bound one needs to assume that $M_A>M_V$ and $F_VG_V>0$, reasonable working assumptions.}
\begin{equation}
\mF_3\Big|^{\mathrm{P{\text -}even}} \,=\,-\displaystyle\frac{v^2}{2M_V^2}\,, \qquad  \qquad
\mF_3\Big|^{\mathrm{P{\text -}even\,\&\,P{\text -}odd}}  \,<\,  -\displaystyle\frac{v^2}{2M_A^2}\,. \label{F3}
\end{equation}
The only $P$-even-operators prediction of (\ref{F3}) is shown by the black curve in the top-right plot of Figure~\ref{plots1}. Adding $P$-odd operators, we have the upper bound of (\ref{F3}), which is represented by the same curve but this time with $M_R=M_A$. Therefore, the whole region below this line (gray area) would be allowed in the most general case.
\item $\mF_4$. In order to get a determination in terms of only resonance masses, one needs to consider a well-behaved two-Goldstones VFF, both WSRs and only $P$-even operators~\cite{PRD},
\begin{equation}
\mF_4\Big|^{\mathrm{1st\,\&\,2nd\,WSRs;\,P{\text -}even}}\,=\, \displaystyle\frac{v^2}{4} \left( \frac{1}{M_V^2}\!-\!\frac{1}{M_A^2}\right)\,.\label{F4}
\end{equation}
We show it in the bottom-left panel in Figure~\ref{plots1}. The upper bound (dark gray curve) is obtained at $M_A\to\infty$. Thus, the theoretically allowed region is the gray area below that curve. The values of $\mF_4$ for some representative axial-vector masses are shown in the red ($M_A=1.2\, M_V$), blue ($M_A=1.1\, M_V$) and orange ($M_A=M_V$) curves again. 
\item $\mF_4+\mF_5$. No constraints are required in this case, since directly from (\ref{integration}) one gets~\cite{PRD} 
\begin{equation}
\mF_4+\mF_5\,=\,\displaystyle\frac{c_d^2}{4M_{S^1_1}^2}\,. \label{F45}
\end{equation}
This clean prediction is shown by the black curve in the bottom-right plot of Figure~\ref{plots1}.
\end{enumerate}

\section{Phenomenology}

\begin{table}[tb]  
\begin{center}
\renewcommand{\arraystretch}{1.0}
\begin{tabular}{|r@{$\,<\,$}c@{$\,<\,$}l|c|c||r@{$\,<\,$}c@{$\,<\,$}l|c|c| }
\hline
\multicolumn{3}{|c|}{LEC} & Ref. & Data & \multicolumn{3}{|c|}{LEC} & Ref. & Data \\ 
\hline \hline
$0.89$ & $\kappa_W$ & $1.13$  & \cite{deBlas:2018tjm}
& LHC  &
$-0.06$&$\mathcal{F}_3$ &$0.20$&\cite{Almeida:2018cld} & LEP \& LHC  \\ \hline 
$-1.02$ & $c_{2V}$ & $2.71$ & \cite{ATLAS:2019dgh} & LHC &
$-0.0006$&$\mathcal{F}_4$&$0.0006$&\cite{Sirunyan:2019der} & LHC  \\ \hline
 $-0.004$  &$ \mathcal{F}_1$& $0.004$  & 
 \cite{Tanabashi:2018oca} & LEP & 
$-0.0010$&$\mathcal{F}_4+\mathcal{F}_5$&$0.0010$ &  \cite{Sirunyan:2019der} & LHC  \\  \hline
\end{tabular}
\caption{{\small
Experimental constraints on purely bosonic EWET LECs, at 95\% CL~\cite{PRD}.}} \label{exp}
\end{center}
\end{table}

Experimental constraints on $\mathcal{F}_{1,3,4,5}$ are reported in Table~\ref{exp}~\cite{PRD} and are shown in the plots of Figure~\ref{plots1} as green areas. While these experimental constraints have been obtained at lower scales, the tree-level predictions of (\ref{integration}) are supposed to apply at a scale around the resonance masses. Thus, it is convenient to estimate the one-loop running uncertainties~\cite{Guo:2015isa} due to these different scales and they are indicated in Figure~\ref{plots1} with the dashed green bands that enlarge the experimentally allowed regions. They depend on $\kappa_W$ and $c_{2V}$, which are also given in Table~\ref{exp}.

The principal result of Figure~\ref{plots1} is that the resonance mass scale is pushed to the TeV range, $M_R \gsim 2\,$TeV, in agreement with Ref.~\cite{ST}, where we estimated theoretically the $S$ and $T$ oblique parameters at NLO. Other minor results are obtained~\cite{PRD}: the $S$ oblique parameter provides the most precise LEC determination ($\mF_1$), implying the constraints $M_{V,A}\gsim 2$~TeV (95\% CL);
the triple gauge couplings give a weaker limit on $\mF_3$, translating into the lower bound $M_{V,A}\gsim 0.5$~TeV (95\% CL);
assuming both WSRs and considering only $P$-even operators, the  bounds on $\mF_4$ constrain the mass of the vector resonance to $M_V\gsim 2$~TeV if $M_A/M_V>1.1$ (95\% CL);
the limit on $\mF_4+\mF_5$ implies that the singlet scalar resonance would have a mass $M_{S_1}\gsim 2$~TeV (95\% CL), taking a  $S_1 WW$ coupling close to the $hWW$ one ($c_d\sim v$). Thus, experimental constraints start already to be competitive and, once new data be available, more precise information will be obtained using this kind of approaches.

\end{document}